\documentclass[prd,twocolumn] {revtex4}
\usepackage[latin1]{inputenc}
\usepackage{graphicx}
\usepackage{epsfig}
\usepackage{bm}
\usepackage{amssymb}
\usepackage{float}
\usepackage{amsmath}
\usepackage{dcolumn}
\usepackage{cancel}
\usepackage{color}
\usepackage{epstopdf}
\usepackage{nccbbb}
\usepackage{lipsum}

\begin{document}

\title {Observational constraints on non-cold dark matter and phenomenological emergent dark energy}

\author{Yan-Hong Yao}
\email{yaoyh29@mail.sysu.edu.cn}

\author{Jun-Chao Wang}

\affiliation{School of Physics and Astronomy, Sun Yat-sen University,
2 Daxue Road, Tangjia, Zhuhai, People's Republic of China}

\author{Xin-He Meng}

\affiliation{School of Physics, Nankai University, Tianjin, 300071, People's Republic of China}

\begin{abstract}
It is well known that there are several long-standing problems implying the discordance of the $\Lambda$CDM model. Although most of the models proposed to resolve these problems assume that dark matter is pressureless, it is still possible that dark matter is not cold, as current observations have not ruled out this possibility yet. Therefore, in this article, we treat the dark matter equation of state parameter as a free parameter, and apply observational data to investigate the non-coldness of dark matter. Impressing by the simplicity of the phenomenological emergent dark energy (PEDE) and its ability to relieve the Hubble tension, we propose the PEDE+$w_{\rm dm}$ model based on PEDE and non-cold dark matter. We then place constraints on this model in light of the Planck 2018 Cosmic Microwave Background (CMB) anisotropies, baryon acoustic oscillation (BAO) measurements, and the Pantheon compilation of Type Ia supernovae. The results indicate a preference for a negative dark matter equation of state parameter at $95\%$ CL for all data sets except CMB alone and CMB+BAO, which suggests that the non-coldness assumption of dark matter worth to be investigated further in order to understand the nature of dark matter. The Hubble tension is alleviated in this scenario compared to the $\Lambda$CDM model, with a significance below 3$\sigma$ level for all data sets except CMB+Pantheon. However, from the analysis based on Bayesian evidence, we clearly see that the data sets favor $\Lambda$CDM over the PEDE+$w_{\rm dm}$ model.
\textbf{}
\end{abstract}

\maketitle

\section{Introduction}
\label{intro}
Although the $\Lambda$ - Cold Dark Matter ($\Lambda$CDM) model has been very successful in fitting a series of astronomical data sets, the nature of dark matter and dark energy, however, has remained mysterious over the last two decades. To exacerbate the situation, such model in fact is theoretical problematic because of two sharp puzzles, i.e. the so-called fine-tuning and coincidence problems. To alleviate these two puzzles, many other alternative scenarios have been
proposed, which can be classified into the categories of dynamical dark energy models and modified gravity models, see for instance Refs.~\cite{Caldwell1998Cosmological,Caldwell1999A,Piazza2004Dilatonic,Padmanabhan2002Accelerated,Bo2006Oscillating,maartens2010brane,esposito2001scalar,capozziello2005reconciling,das2006curvature}.
Furthermore, Owing to the higher level of accuracy achieved in astronomical observations, the tensions in the cosmological parameters have been quite serious in recent years~\cite{di2021realm,perivolaropoulos2022challenges,schoneberg2022h0,abdalla2022cosmology}, this has prompted numerous astronomers to put forth various models in order to alleviate these tension. Among these solutions, early dark energy~\cite{reeves2023restoring,vagnozzi2021consistency,poulin2019early,agrawal2019rock,smith2020oscillating,lin2019acoustic,niedermann2021new,freese2021chain,ye2020hubble,akarsu2020graduated,braglia2020unified}, late dark energy~\cite{vagnozzi2018constraints,visinelli2019revisiting,huang2016dark,vagnozzi2020new,martinelli2019cmb,alestas2020h,d2021limits,yang2019observational,alestas2021late}, dark radiation~\cite{battye2014evidence,Zhang2014Neutrinos,zhang2015sterile,feng2018searching,zhao2018measuring,choudhury2019constraining} and interacting dark energy~\cite{yao2023can,yao2021relieve,yao2020new1,kumar2016probing,kumar2017echo,kumar2019dark,nunes2022new,yang2018tale,di2017can,yang2018interacting,di2020nonminimal,di2020interacting,cheng2020testing,lucca2020tensions,gomez2020update,yang2019dark,yang2020dynamical} have been researched extensively. If one carefully examines the existing literature, it is found that there is a common practice in constructing cosmological models, which is to assume that the dark matter equation of state is zero, that is, to assume that dark matter is pressureless. Of course, this assumption is not made arbitrarily, it is supported by a large body of cosmological observations, including observations of large-scale structures. However, since the nature of dark matter is not known yet, by treating the dark matter equation of state parameter as a free parameter, we can employ observational data to verify whether it is zero. This approach has already been used by several works to test the non-coldness of dark matter with multiple types of measurement data~\cite{mueller2005cosmological,kumar2014observational,armendariz2014cold,kopp2018dark,kumar2019testing,ilic2021dark,pan2022iwdm}.

In this paper, we focus on the phenomenological emergent dark energy (PEDE) and non-cold dark matter with equation of state that varies freely, the PEDE model was proposed by the authors of Ref.~\cite{li2019simple} for the purpose to resolve the Hubble tension and subsequently be extended to a generalized parameterization form that can include both cosmological constant and the PEDE model~\cite{li2020evidence} (also see~\cite{yang2021generalized,hernandez2020generalized}). The PEDE model has exactly the same number of free parameters as the spatially flat $\Lambda$CDM model. This scenario was investigated further in Ref.~\cite{pan2020reconciling} considering its complete evolution including background as well as perturbations and obtained similar results concerning the $H_0$ tension with the recent observational data sets.  In order to check how the extra degrees of freedom in terms of the neutrino properties could affect the constraints on the Hubble constant, the PEDE+$M_{\nu}+N_{\rm eff}$ model ($M_{\nu}$ is the total neutrino mass and $N_{\rm eff}$ is the effective number of neutrino species) is considered by the authors of Ref.~\cite{yang2021emergent}, and with the full data sets used in Ref.~\cite{yang2021emergent}, they found, at $95\%$ level, $M_{\nu}\sim0.21_{-0.14}^{+0.15}$ eV and $N_{\rm eff}\sim3.03\pm0.32$, i.e. an indication for a non-zero neutrino mass with a significance above 2$\sigma$. This work motivates us to propose the PEDE+$w_{\rm dm}$ model in this paper to test whether the non-zero dark matter equation of state parameter is preferred by the current observations with a significance above 2$\sigma$.

The rest of this paper is organized as follows. In section~\ref{sec:1}, we describe the key equations of the PEDE+$w_{\rm dm}$ model. In section~\ref{sec:2}, we
describe the observational data sets and the statistical methodology. In section~\ref{sec:3}, we describe the observational
constraints and the implications of the PEDE+$w_{\rm dm}$ model. In the last section, we make a brief conclusion with this paper.
\section{Introducing the PEDE+$w_{\rm dm}$ model}
\label{sec:1}
We consider a spatially flat, homogeneous and isotropic space-time which is described by the spatially flat Friedmann-Robertson-Walker (FRW) metric, we also consider that the gravitational sector of the universe is well described by general relativity in which matter is minimally coupled to gravity. In addition to that, we further assume
that none of the fluids are interacting non-gravitationally with each other, and the universe is comprised of radiation, baryons, non-cold dark matter and PEDE. Therefore, one can write down the Hubble equation as
\begin{equation}
\frac{H^2}{H_0^2}
 =\Omega_{\rm r0}(1+z)^{4}+\Omega_{\rm dm0}(1+z)^{3(1+w_{\rm dm})}+\Omega_{\rm b0}(1+z)^{3}+\Omega_{\rm de}(z),\hspace{1cm}
\end{equation}
where $H$ is the Hubble parameter, $\Omega_{\rm r0}$ is the density parameter for radiation, $\Omega_{\rm dm0}$ is the density parameter for non-cold dark matter, $\Omega_{\rm b0}$ is the density parameter for baryons, and $\Omega_{\rm de}(z)$ is the PEDE density parameter which is parameterized in the following form~\cite{li2019simple,pan2020reconciling}:
\begin{equation}\label{eq:1}
\Omega_{\rm de}(z)=\Omega_{\rm de0}[1-\tanh(\log_{10}(1+z))],
\end{equation}
where $\Omega_{\rm de0}=1-\Omega_{\rm r0}-\Omega_{\rm dm0}-\Omega_{\rm b0}$ and $1+z=a_0a^{-1}=a^{-1}$ (without losing any generality we set the current
value of the scale factor $a_0$ to be unity). Since there is no interaction between any two fluids under consideration, therefore, the conservation equation for dark energy reads
\begin{equation}\label{}
  \dot{\rho}_{\rm de}(z)+3H(1+w_{\rm de}(z))\rho_{\rm de}(z) =0,
\end{equation}
here, an over dot stands for the cosmic time derivative, from the above equation we can derive the equation of state for dark energy as
\begin{equation}\label{}
  w_{\rm de}(z)=-1+\frac{1}{1+z}\times\frac{d\ln\Omega_{\rm de}(z)}{dz},
\end{equation}
substituting Eq.(\ref{eq:1}) in it, we obtain the equation of state for PEDE as follows~\cite{li2019simple,pan2020reconciling}
\begin{equation}\label{}
  w_{\rm de}(z)=-1-\frac{1}{3\ln10}\times[1+\tanh(\log_{10}(1+z))],
\end{equation}
it can be inferred from the above equation that PEDE's equation of state has an interesting symmetrical feature, specifically speaking, for far past, i.e. $z\rightarrow\infty$, one obtains $w_{\rm de}\rightarrow-1-\frac{2}{3\ln10}$, and for far future, i.e. $z\rightarrow-1$, one finds $w_{\rm de}\rightarrow-1$, as for the present time, namely $z=0$, we see $w_{\rm de}=-1-\frac{1}{3\ln10}$, which is a phantom dark energy equation of state. As described briefly in the Ref.~\cite{li2019simple}, the pivot point of transition of the PEDE equation of state can be considered to be the redshift of matter-dark energy densities equality.

In the conformal Newtonian gauge, the perturbed FRW metric takes the form
\begin{equation}\label{}
  ds^2=a^2(\tau)[-(1+2\psi)d\tau^2+(1-2\phi)d\vec{r}^2],
\end{equation}
where $\psi$ and $\phi$ are the metric potentials and $\vec{r}$ represents the three spatial coordinates, from the first order perturbed part of the conserved stress-energy momentum
tensor, one obtains the following continuity and Euler equations (in the Fourier space) for non-cold dark matter and PEDE~\cite{kumar2019testing}.
\begin{widetext}
\begin{equation}
 \delta^{\prime}_{\rm dm}= -(1+w_{\rm dm}) \left(\theta_{\rm dm} - 3 \phi^{\prime} \right)
 -3 \mathcal{H} \delta_{\rm dm} (c^2_{\rm s,dm} - w_{\rm dm}) - 9 (1+w_{\rm dm})(c^2_{\rm s,dm} - c^2_{\rm a,dm})\mathcal{H}^2 \frac{\theta_{\rm dm}}{k^2},
\end{equation}
\begin{equation}
\theta^{\prime}_{\rm dm}=-(1-3 c^2_{\rm s,dm}) \mathcal{H} \theta_{\rm dm}  + \frac{c^2_{\rm s,dm}}{1+w_{\rm dm}}k^2 \delta_{\rm dm} + k^2\psi.
\end{equation}
\begin{equation}
 \delta^{\prime}_{\rm ds}= -(1+w_{\rm ds}) \left(\theta_{\rm ds} - 3 \phi^{\prime} \right)
 -3 \mathcal{H} \delta_{\rm ds} (c^2_{\rm s,ds} - w_{\rm ds}) - 9 (1+w_{\rm ds})(c^2_{\rm s,ds} - c^2_{\rm a,ds})\mathcal{H}^2 \frac{\theta_{\rm ds}}{k^2},
\end{equation}
\begin{equation}
\theta^{\prime}_{\rm ds}=-(1-3 c^2_{\rm s,ds}) \mathcal{H} \theta_{\rm ds}  + \frac{c^2_{\rm s,ds}}{1+w_{\rm ds}}k^2 \delta_{\rm ds} + k^2\psi.
\end{equation}
\end{widetext}
Here, a prime stands for the conformal time derivative, $\mathcal{H}$ is the conformal Hubble parameter, and k is magnitude of the wavevector $\vec{k}$. Further, $\delta_{\rm dm} (\delta_{\rm de})$ and $\theta_{\rm dm}(\theta_{\rm de})$ are the relative density and velocity divergence perturbations of dark matter (dark energy), $c_{\rm s,dm}(c_{\rm s,de})$ and $c_{\rm a,dm}(c_{\rm a,de})$ denote the sound speed and adiabatic sound speed of dark matter (dark energy). The sound speed of dark sectors describe their micro-scale properties and need to be provided independently, in this article, we consider $c^2_{\rm s,dm}=0$ and $c^2_{\rm s,de}=1$. Having presented the equations above, the background and perturbation dynamics of the PEDE+$w_{\rm dm}$ model is clearly understood.

\begin{figure*}
\begin{minipage}{0.45\linewidth}
  \centerline{\includegraphics[width=1\textwidth]{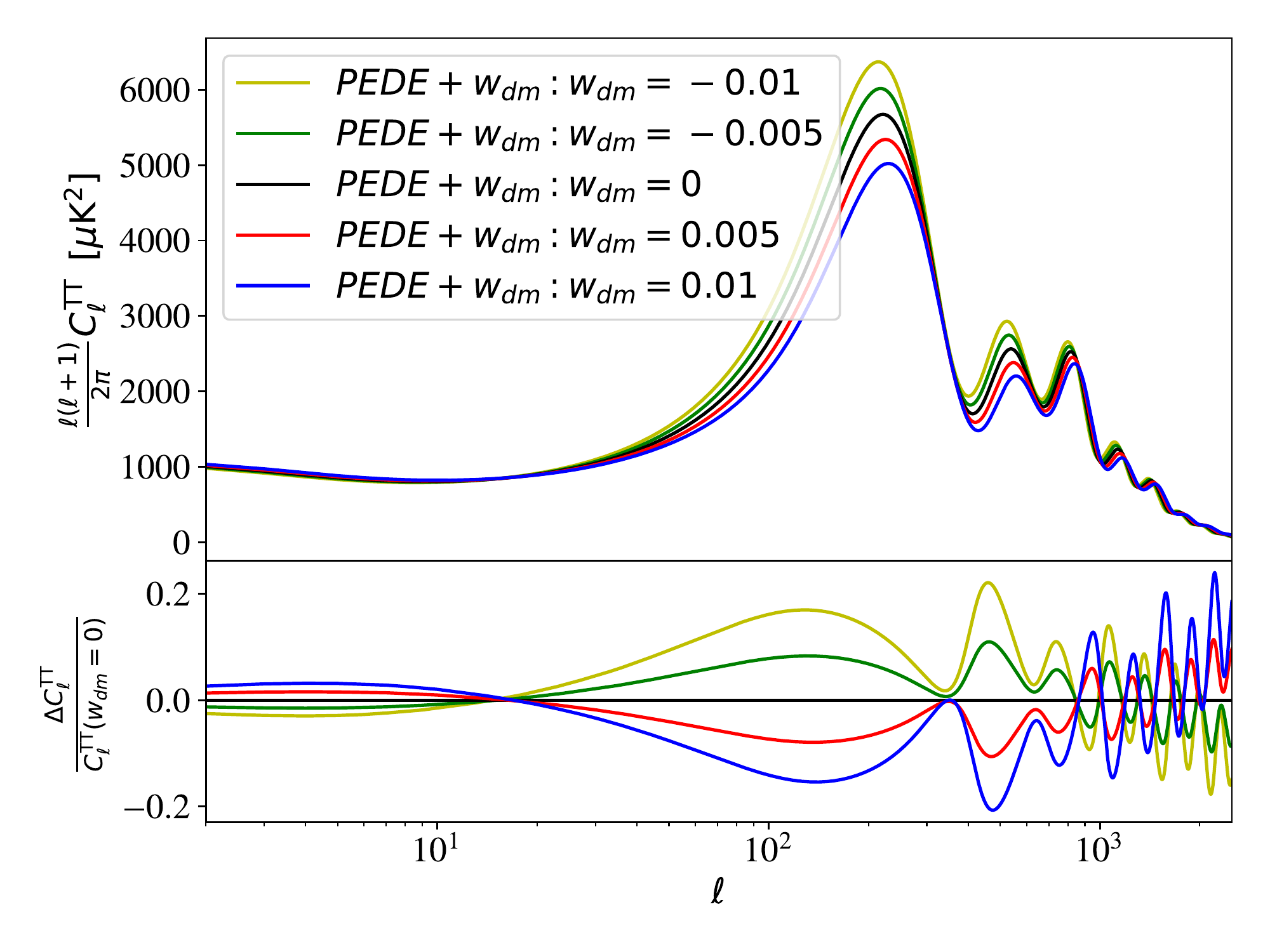}}
  \label{a}
\end{minipage}
\begin{minipage}{0.45\linewidth}
  \centerline{\includegraphics[width=1\textwidth]{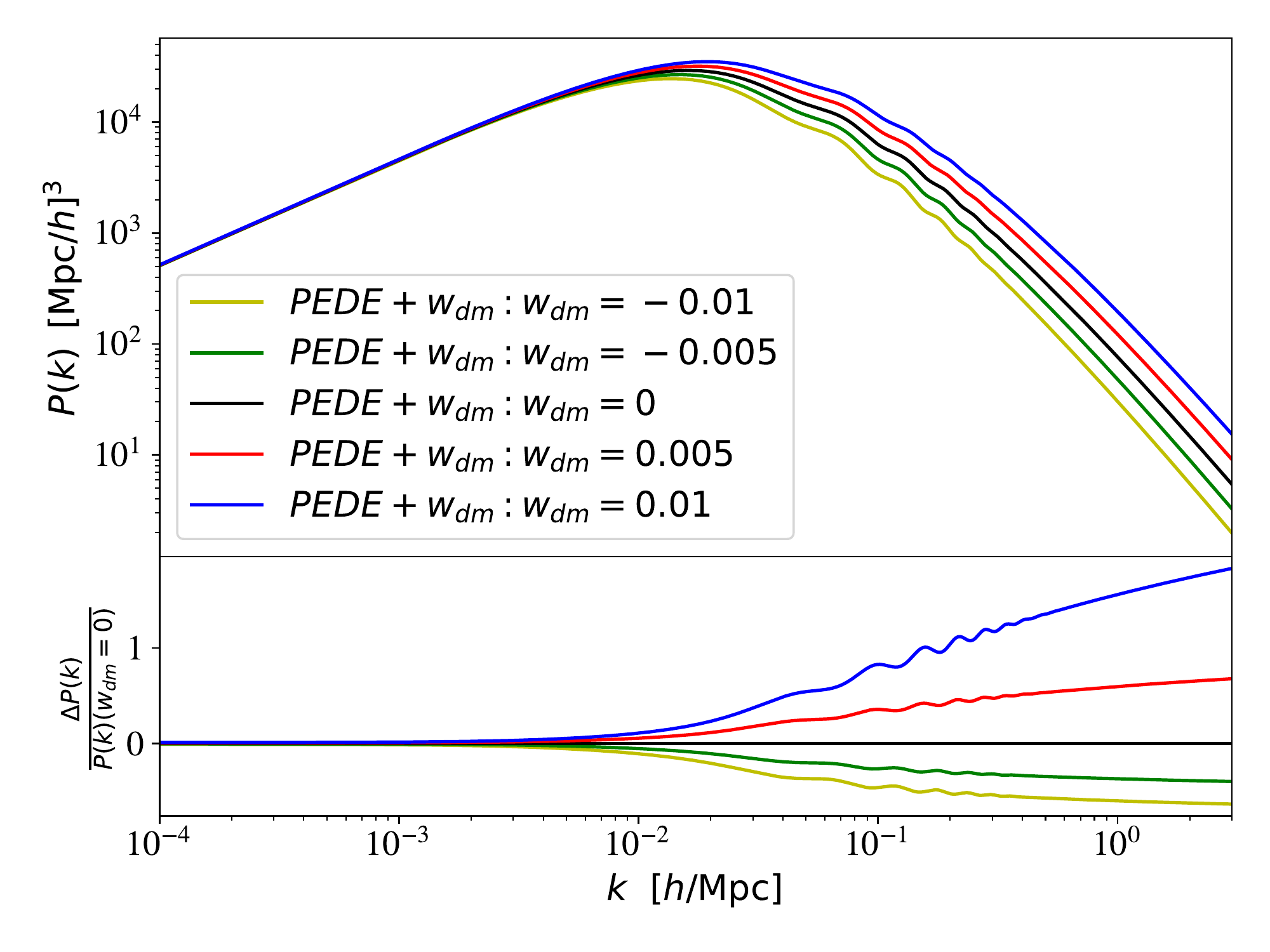}}
\end{minipage}
\caption{The CMB TT and matter power spectra for different values of parameter $w_{dm}$, where other relevant model parameters are fixed to their mean values extracted from CMB+BAO+Pantheon data analysis.}
\label{fig:2}
\end{figure*}

In the end of this section we present some analysis concerning the impacts of the PEDE+$w_{\rm dm}$ model on the CMB TT and matter power spectra for different values of $w_{\rm dm}$. In Fig.\ref{fig:2} we plot the CMB TT and the matter power spectra by setting $w_{\rm dm}=-0.01$, $-0.005$, $0$, $0.005$, $0.01$ and fixing six other parameters to their mean values extracted from CMB+BAO+Pantheon data analysis.

Focus on the CMB TT power spectrum, we see that the amplitudes of the acoustic peaks of CMB predicted by PEDE+$w_{\rm dm}$ with negative values of parameter $w_{\rm dm}$ are increased and the positions of these acoustic peaks are moved to the left side, this is because negative values of parameter $w_{\rm dm}$ will increase the equality time of matter and radiation when other model parameters are fixed. On the large scale where $l<10$, the curves are depressed when the values of parameter $w_{\rm dm}$ are negative due to the integrated Sachs-Wolfe effect. In addition, when the values of parameter $w_{\rm dm}$ become positive while keeping the values of other relevant cosmological parameters unchanged, the resulting effects will be opposite to those caused by considering negative values of $w_{\rm dm}$.

For the matter power spectrum, we can see that negative values of $w_{\rm dm}$ decrease the matter power spectrum and positive values of $w_{\rm dm}$ increase the matter power spectrum, this is because the former situation results in a delay of matter and radiation equality, while the latter situation leads to an advance of matter and radiation equality.
\section{Data sets and methodology}
\label{sec:2}
To extract the free parameters of the PEDE+$w_{\rm dm}$ model, we use the recent observational data sets described below.

\textbf{Cosmic Microwave Background (CMB)}: we make use of the Planck 2018~\cite{aghanim2020planck1,aghanim2020planck2} CMB data, more specifically, we use the CMB temperature
and polarization angular power spectra plikTTTEEE+lowl+lowE. In addition to that, Planck 2018 CMB lensing reconstruction likelihood~\cite{aghanim2020planck3} also be included.

\textbf{Baryon acoustic oscillations(BAO)}:BAO distance measurements from
several astronomical surveys including 6dFGS~\cite{Beutler2011The}, SDSS-MGS~\cite{ross2015clustering} and BOSS DR12~\cite{alam2017clustering} are considered.

\textbf{Pantheon}: We also include the Pantheon catalogue of the Supernovae Type Ia, comprising 1048 data
points in the redshift region $z\in[0.01, 2.3]$.

We also present the latest local measurement of the Hubble constant yielding $H_0=73.04\pm1.04$ obtained by SH0ES collaboration~\cite{riess2022comprehensive} here, and denote it as R22 hereafter. (Although we will not use this prior to constrain the model, we will use it to calculate the Hubble tensions in the PEDE+$w_{\rm dm}$ model concerning different data sets.)
To constrain the PEDE+$w_{\rm dm}$ model, we run a Markov Chain Monte Carlo (MCMC) using the public code MontePython-v3~\cite{audren2013conservative,brinckmann2019montepython} and a modified version of the CLASS code~\cite{lesgourgues2011cosmic,blas2011cosmic}. We perform the analysis with a Metropolis-Hasting algorithm and consider chains to
be converged using the Gelman-Rubin~\cite{gelman1992inference} criterion $R-1<0.03$. In Table \ref{tab:priors} we display the priors on the free parameters of
PEDE+$w_{\rm dm}$, that are, the baryon density $\omega_{\rm b}$, the non-cold dark matter density $\omega_{\rm dm}$, the ratio of the sound horizon to the angular diameter
distance $\theta_s$, the amplitude and the spectral index of the primordial scalar perturbations $A_s$ and $n_s$, the optical depth $\tau_{\rm reio}$, and the equation of state of the non-cold dark matter $w_{\rm dm}$. Finally, we use the MCEvidence code~\cite{heavens2017no,heavens2017marginal} to compute the Bayesian Evidence for all the models considered in this work, this can be done by only using the MCMC chains which are used to extract the cosmological parameters. The Bayesian evidence, $B$, is given by the integral of the prior $\pi$ times the likelihood $L$ over the entire parameter space of the model
\begin{equation}\label{}
  B=\int L\pi d\theta,
\end{equation}
models with larger Bayesian evidence are considered to be more supported by the data. In practice, one often uses the logarithm of the Bayes factor $B_{ij}$, i.e. logarithm of the evidence ratio for model $i$ and $j$, to perform model selection. In our situation, after we obtain $\ln B_{ij}$ (where $i$=PEDE, or PEDE+$w_{\rm dm}$, $j$=$\Lambda$CDM) with the help of the MCEvidence code, we can use the revised Jeffreys' scale~\cite{kass1995bayes}, i.e. empirical scale for evaluating the strength of evidence
when comparing two models, shown in Table~\ref{tab:scale} to quantify the strength of evidence of the PEDE model and the PEDE+$w_{\rm dm}$ model with respect to the $\Lambda$CDM model.
\begin{table}
\begin{center}
\begin{tabular}{c c}
\hline
Parameter & Prior\\
\hline
\hline
$100 \omega_{\rm b}$ & [0.8, 2.4]\\
$\omega_{\rm dm}$ & [0.01, 0.99] \\
$100\theta_s$ & [0.5, 2.0] \\
$\ln[10^{10}A_{s }]$ & [2.7, 4.0]\\
$n_s$ & [0.9, 1.1] \\
$\tau_{\rm reio}$  & [0.01,  0.8] \\
$w_{\rm dm}$  & [$-$0.1, 0.1] \\
\hline
\hline
\end{tabular}
\end{center}
\caption{Uniform priors on the free parameters of the PEDE+$w_{\rm dm}$ model.} \label{tab:priors}
\end{table}

\begin{table}
\begin{center}
\begin{tabular}{c c }
               \hline
	            \hline
 $ \ln B_{ij}$ &  Strength of evidence for model $M_{i}$
\\ \hline
$ 0\leq\ln B_{ij}<1$    &Weak
                     \\
$1\leq \ln B_{ij}<3$&  Positive
                       \\
$3\leq \ln B_{ij}<5$&  Strong
                       \\
$ \ln B_{ij}\geq5$&  Very strong
                     \\
\hline
\hline
\end{tabular}
\end{center}
\caption{Revised Jeffreys' scale which quantifies the strength of evidence of model $M_{i}$ with respect to model $M_{j}$.}
\label{tab:scale}
\end{table}

\section{Results and discussion}
\label{sec:3}
\begin{table*}[!t]
	\renewcommand\arraystretch{1.1}
	\setlength{\tabcolsep}{0.1mm}{
		{\begin{tabular}{c c c c c }
               \hline
	            \hline
 Parameters & ${\rm CMB}$ & ${\rm CMB+BAO}$&${\rm CMB+Pantheon}$&${\rm CMB+BAO+Pantheon}$
\\ \hline
$100\omega_{\rm b}$    &${2.239^{+0.015}_{-0.015}}^{+0.029}_{-0.030}$ &  $ {2.227^{+0.013}_{-0.013}}^{+0.026}_{-0.026}  $ & $  {2.226^{+0.014}_{-0.014}}^{+0.027}_{-0.027} $&${2.221^{+0.013}_{-0.013}}^{+0.026}_{-0.026}   $
                     \\
$\omega_{\rm dm}$   &    ${0.1198^{+0.0012}_{-0.0012}}^{+0.0024}_{-0.0024}$ &  ${0.1213^{+0.0009}_{-0.0009}}^{+0.0018}_{-0.0018} $& $  {0.1214^{+0.0011}_{-0.0011}}^{+0.0023}_{-0.0023}$ &${0.1221^{+0.0009}_{-0.0009}}^{+0.0016}_{-0.0016}$
                       \\
 $100\theta_s$          &    $ {1.04189^{+0.00030}_{-0.00030}}^{+0.00059}_{-0.00059} $ & ${1.04175^{+0.00030}_{-0.00030}}^{+0.00058}_{-0.00060} $& $ {1.04177^{+0.00030}_{-0.00030}}^{+0.00057}_{-0.00060} $&${1.04172^{+0.00028}_{-0.00028}}^{+0.00057}_{-0.00053}$
                       \\
$\ln[10^{10}A_{s }]$         &  ${3.043^{+0.015}_{-0.015}}^{+0.029}_{-0.028}$ &  $ {3.039^{+0.014}_{-0.014}}^{+0.027}_{-0.028} $& ${3.037^{+0.014}_{-0.014}}^{+0.028}_{-0.029}     $ &${3.036^{+0.014}_{-0.014}}^{+0.026}_{-0.027} $
                     \\
$n_s$         &  ${0.9655^{+0.0042}_{-0.0042}}^{+0.0080}_{-0.0084} $ & ${0.9619^{+0.0036}_{-0.0036}}^{+0.0071}_{-0.0068} $& $  {0.9617^{+0.0040}_{-0.0040}}^{+0.0079}_{-0.0081}  $&$ {0.9603^{+0.0035}_{-0.0035}}^{+0.0068}_{-0.0069} $
                     \\
$\tau_{\rm reio}$  &  $ {0.054^{+0.008}_{-0.008}}^{+0.015}_{-0.014}   $     &  $ {0.050^{+0.007}_{-0.007}}^{+0.014}_{-0.014} $      &${0.050^{+0.007}_{-0.007}}^{+0.014}_{-0.015}$ &$ {0.048^{+0.007}_{-0.007}}^{+0.013}_{-0.014}  $
                \\
$\Omega_{\rm m}$  &  ${0.270^{+0.007}_{-0.007}}^{+0.015}_{-0.014}$      &  $ {0.280^{+0.006}_{-0.006}}^{+0.011}_{-0.011}  $         &${0.280^{+0.007}_{-0.007}}^{+0.014}_{-0.013}$ &$ {0.285^{+0.005}_{-0.005}}^{+0.010}_{-0.010}$
                     \\
$ H_0  $             &  ${72.5^{+0.70}_{-0.70}}^{+1.4}_{-1.4} $      &  ${71.7^{+0.51}_{-0.51}}^{+1.0}_{-1.0}   $         &${71.6^{+0.63}_{-0.63}}^{+1.2}_{-1.3}$ &$ {71.21^{+0.47}_{-0.47}}^{+0.91}_{-0.91} $
                          \\
$\sigma_8$ & $  {0.856^{+0.006}_{-0.006}}^{+0.012}_{-0.012}      $&$  {0.858^{+0.006}_{-0.006}}^{+0.012}_{-0.012}  $ & ${0.857^{+0.006}_{-0.006}}^{+0.012}_{-0.013}      $&  ${0.859^{+0.006}_{-0.006}}^{+0.012}_{-0.012}      $
\\
\hline
\hline
\end{tabular}
\caption{The mean values and 1, 2 $\sigma$ errors of the parameters of the PEDE model for CMB, CMB+BAO, CMB+Pantheon, and CMB+BAO+Pantheon data sets.}
\label{tab:1}}}
\end{table*}
\begin{table*}[!t]
	\renewcommand\arraystretch{1.1}
	\setlength{\tabcolsep}{0.05mm}{
		{\begin{tabular}{c c c c c }
               \hline
	            \hline
 Parameters & ${\rm CMB}$ & ${\rm CMB+BAO}$&${\rm CMB+Pantheon}$&${\rm CMB+BAO+Pantheon}$
\\ \hline
$100\omega_{\rm b}$    &${2.241^{+0.016}_{-0.016}}^{+0.032}_{-0.030}$ &  $ {2.241^{+0.015}_{-0.015}}^{+0.030}_{-0.030}  $ & $  {2.244^{+0.015}_{-0.015}}^{+0.031}_{-0.030} $&${2.242^{+0.015}_{-0.015}}^{+0.030}_{-0.029}   $
                     \\
$\omega_{\rm dm}$   &    ${0.1228^{+0.0062}_{-0.0027}}^{+0.0074}_{-0.0092}$ &  ${0.1225^{+0.0011}_{-0.0011}}^{+0.0021}_{-0.0021} $& $  {0.1267^{+0.0021}_{-0.0017}}^{+0.0037}_{-0.0037}$ &${0.1237^{+0.0010}_{-0.0010}}^{+0.0020}_{-0.0020}$
                       \\
 $100\theta_s$          &    $ {1.04185^{+0.00033}_{-0.00033}}^{+0.00062}_{-0.00064} $ & ${1.04185^{+0.00029}_{-0.00029}}^{+0.00058}_{-0.00056} $& $ {1.04176^{+0.00031}_{-0.00031}}^{+0.00061}_{-0.00061} $&${1.04185^{+0.00030}_{-0.00030}}^{+0.00057}_{-0.00058}$
                       \\
$\ln[10^{10}A_{s }]$         &  ${3.040^{+0.015}_{-0.015}}^{+0.032}_{-0.029}$ &  $ {3.040^{+0.014}_{-0.014}}^{+0.027}_{-0.028} $& ${3.041^{+0.012}_{-0.016}}^{+0.031}_{-0.029}     $ &${3.040^{+0.014}_{-0.014}}^{+0.028}_{-0.028} $
                     \\
$n_s$         &  ${0.964^{+0.005}_{-0.005}}^{+0.010}_{-0.010} $ & ${0.9639^{+0.0038}_{-0.0038}}^{+0.0077}_{-0.0071} $& $  {0.9612^{+0.0045}_{-0.0045}}^{+0.0085}_{-0.0080}  $&$ {0.9630^{+0.0037}_{-0.0037}}^{+0.0073}_{-0.0071} $
                     \\
$\tau_{\rm reio}$  &  $ {0.053^{+0.008}_{-0.008}}^{+0.016}_{-0.015}   $     &  $ {0.053^{+0.007}_{-0.007}}^{+0.015}_{-0.014} $      &${0.054^{+0.006}_{-0.008}}^{+0.016}_{-0.015}$ &$ {0.053^{+0.007}_{-0.007}}^{+0.014}_{-0.013}  $
                \\
$ w_{\rm dm}$   &  $ {-0.0009^{+0.0011}_{-0.0019}}^{+0.0032}_{-0.0026} $      &  $  {-0.00090^{+0.00047}_{-0.00047}}^{+0.00092}_{-0.00092}  $         &${-0.0023^{+0.0006}_{-0.0007}}^{+0.0013}_{-0.0013}$ &${-0.00130^{+0.00042}_{-0.00042}}^{+0.00080}_{-0.00085}  $
                 \\
$\Omega_{\rm m}$  &  ${0.298^{+0.051}_{-0.031}}^{+0.068}_{-0.076}$      &  $ {0.294^{+0.009}_{-0.009}}^{+0.018}_{-0.018}  $         &${0.333^{+0.020}_{-0.020}}^{+0.036}_{-0.035}$ &$ {0.304^{+0.009}_{-0.009}}^{+0.017}_{-0.016}$
                     \\
$ H_0  $             &  ${70.2^{+2.2}_{-4.9}}^{+7.6}_{-6.0} $      &  ${70.3^{+0.86}_{-0.86}}^{+1.7}_{-1.6}   $         &${67.0^{+1.2}_{-1.6}}^{+2.7}_{-2.7}$ &$ {69.4^{+0.76}_{-0.76}}^{+1.5}_{-1.5} $
                          \\
$ \sigma_8$       &   $ {0.833^{+0.028}_{-0.050}}^{+0.084}_{-0.068}        $   &  $ {0.833^{+0.014}_{-0.014}}^{+0.027}_{-0.027}       $   &  ${0.801^{+0.015}_{-0.019}}^{+0.033}_{-0.034}      $  & ${0.823^{+0.013}_{-0.013}}^{+0.025}_{-0.024}$
\\
\hline
\hline
\end{tabular}
\caption{The mean values and 1, 2 $\sigma$ errors of the parameters of the PEDE+$w_{\rm dm}$ model for CMB, CMB+BAO, CMB+Pantheon, and CMB+BAO+Pantheon data sets.}
\label{tab:2}}}
\end{table*}
\begin{figure*}
	\centering
	\includegraphics[scale=0.5]{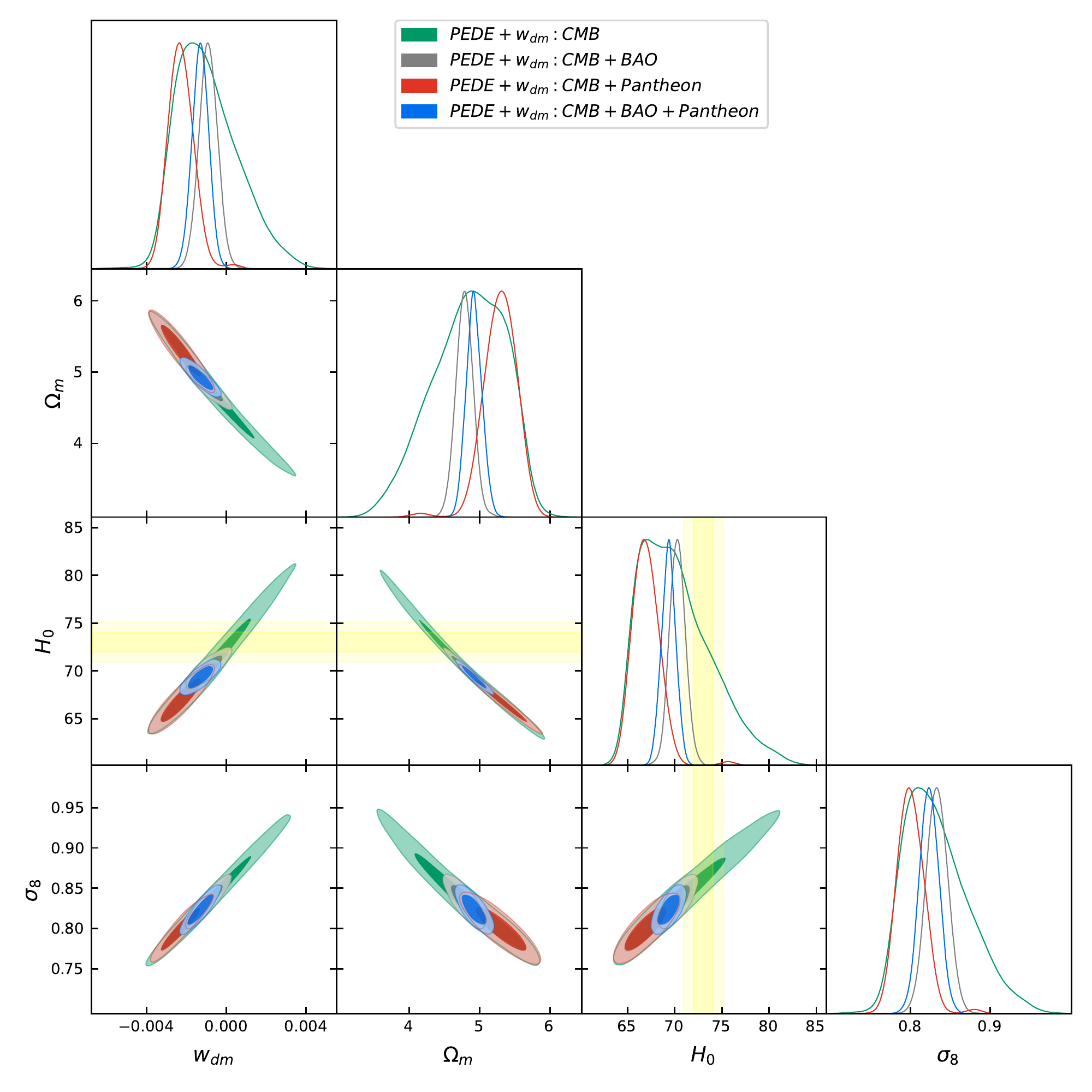}
	\caption{one dimensional posterior distributions and two dimensional joint contours at 68\% and 95\% CL for the most relevant parameters of the PEDE+$w_{\rm dm}$ model using CMB, CMB+BAO, CMB+Pantheon, and CMB+BAO+Pantheon observational data sets. The R22 is shown in yellow. }
\label{fig:1}
\end{figure*}
\begin{table*}[!t]
	\renewcommand\arraystretch{1.1}
	\setlength{\tabcolsep}{0.6mm}{
		{\begin{tabular}{c c c }
               \hline
	            \hline
 Model &  Datasets & $ \ln B_{ij}$
\\ \hline
PEDE    &CMB &0.93
                     \\
PEDE&   CMB+BAO &$-2.32$
                       \\
PEDE&  CMB+Pantheon &$-5.7$
                       \\
PEDE&  CMB+BAO+Pantheon& $-8.02$
                     \\

                      \hline

${\rm PEDE}+w_{\rm dm}$   &CMB &$-0.76$
                     \\
${\rm PEDE}+w_{\rm dm}$&   CMB+BAO &$-3.37$
                       \\
${\rm PEDE}+w_{\rm dm}$&  CMB+Pantheon &$-2.03$
                       \\
${\rm PEDE}+w_{\rm dm}$&  CMB+BAO+Pantheon& $-6.45$
                     \\

\hline
\hline
\end{tabular}
\caption{Summary of the $\ln B_{ij}$ values quantifying the evidence of fit of the PEDE model and the PEDE+$w_{\rm dm}$ model with respect to the $\Lambda$CDM model under the CMB, CMB+BAO, CMB+Pantheon, and CMB+BAO+Pantheon data sets. One note that a negative value of $\ln B_{ij}$ indicates the PEDE model or the PEDE+$w_{\rm dm}$ model is less supported compared to the base model $\Lambda$CDM.}
\label{tab:3}}}
\end{table*}

In Tab.~\ref{tab:2} and Fig.~\ref{fig:1}, we present the constraints on the PEDE+$w_{\rm dm}$ model for CMB, CMB+BAO, CMB+Pantheon, and CMB+BAO+Pantheon data sets. For comparison, we also present the mean values and 1, 2 $\sigma$ errors of the parameters of the PEDE model for the same data sets in the Tab.~\ref{tab:1}.

We begin by discussing the fitting results of the PEDE+$w_{\rm dm}$ model for CMB alone and then gradually investigate the effects of other probes by adding them to CMB. For CMB alone, we find that there is no evidence supporting a non-zero dark matter parameter $w_{\rm dm}$ at $68\%$ CL ($w_{\rm dm}=-0.0009^{+0.0011}_{-0.0019}$ at $68\%$ CL). Nevertheless, some differences in other relevant parameters between the PEDE model and the PEDE+$w_{\rm dm}$ model is presented due to the small but not negligible negative mean value of $w_{\rm dm}$. Firstly, since the mean value of $w_{\rm dm}$ is shifted towards negative, the Hubble constant $H_0$ in the PEDE+$w_{\rm dm}$ model is reduced to $ H_0=70.2^{+2.2}_{-4.9}$ at $68\%$ CL compared to its counterpart in the PEDE model ($H_0=72.5^{+0.70}_{-0.70}$ at $68\%$ CL), due to the positive correlation between $w_{\rm dm}$ and $H_0$. As a results, the $H_0$ tension with R22 increases from 0.4$\sigma$ in the PEDE model to 1.2$\sigma$ in the PEDE+$w_{\rm dm}$ model. Secondly, the positive correlation between $w_{\rm dm}$ and $\sigma_8$ shown in Fig.~\ref{fig:1} results in an decrease in the value of parameter $\sigma_8$ ($\sigma_8=0.833^{+0.028}_{-0.050}$ at $68\%$ CL) in the PEDE+$w_{\rm dm}$ model compared to its counterpart in the PEDE model ($\sigma_8=0.856^{+0.006}_{-0.006}$ at $68\%$ CL), this is consistent with the conclusion that negative values of $w_{\rm dm}$ decrease the matter power spectrum drawn by us in the previous section. Finally, because of the negative correlation between $w_{\rm dm}$ and $\Omega_m$, the matter density parameter $\Omega_m$ increases from $\Omega_m=0.270^{+0.007}_{-0.007}$ at $68\%$ CL in the PEDE model to $\Omega_m=0.298^{+0.051}_{-0.031}$ at $68\%$ CL in the PEDE+$w_{\rm dm}$ model.

When BAO data are added to CMB, we observe some changes in the fitting results. In particular, we find an indication for a negative dark matter equation of state parameter at $68\%$ CL ($w_{\rm dm}=-0.00090^{+0.00047}_{-0.00047}$ at $68\%$ CL). In addition, we obtain a higher $H_0$ ($H_0=70.3^{+0.86}_{-0.86}$ at $68\%$ CL) compared to the CMB analysis, nevertheless, the $H_0$ tension with R22 is aggravated by the decrease of the 1$\sigma$ error of the $H_0$ parameter, to be more specific, it has increased from 1.2$\sigma$ to 2.0$\sigma$.

The inclusion of Pantheon to CMB, shifts the mean value of the parameter $w_{\rm dm}$ more far away from zero compared to the previous case concerning the CMB+BAO data set, with an indication for a negative dark matter equation of state parameter at $95\%$ CL ($w_{\rm dm}=-0.0023^{+0.0013}_{-0.0013}$ at $95\%$ CL). Comparing with the CMB+BAO analysis, we obtain a lower Hubble constant ($H_0=67.0^{+1.2}_{-1.6}$ at $68\%$ CL), aggravating the $H_0$ tension. As a result, the $H_0$ tension with R22 is raised to 3.8$\sigma$. We attribute this outcome to the preference for a higher $\Omega_{\rm m}$ value when Pantheon data set is included and the negative correlation between $H_0$ and $\Omega_{\rm m}$.

For the CMB+BAO+Pantheon data set, we still find a preference for a negative dark matter equation of state parameter at $95\%$ CL ($w_{\rm dm}=-0.00130^{+0.00080}_{-0.00085}$ at $95\%$ CL). Comparing with the previous case which concerning the CMB+Pantheon data set, we find a higher Hubble constant ($H_0=69.4^{+0.76}_{-0.76}$ at $68\%$ CL), relieving the Hubble tension with R22 to 2.8$\sigma$. One notes that the mean values of cosmological parameters extracted from the CMB+BAO+Pantheon data set are between their counterparts constrained from the CMB+BAO and the CMB+Pantheon data sets.

Finally, we present the $\ln B_{ij}$ values quantifying the evidence of fit of the PEDE model and the PEDE+$w_{\rm dm}$ model with respect to the $\Lambda$CDM model under the data sets considered in this work in Tab.~\ref{tab:3}. Recall the Revised Jeffreys' scale shown in Tab.~\ref{tab:scale}, we find that although PEDE+$w_{\rm dm}$ performs better than $\Lambda$CDM when it comes to alleviating tensions in $H_0$ for most data sets, the Bayesian evidence shows that all the data sets considered in this work favor $\Lambda$CDM more. More specifically, there is a weak evidence showing the data sets CMB favor $\Lambda$CDM over PEDE+$w_{\rm dm}$, a positive evidence showing the data set CMB+Pantheon favor $\Lambda$CDM over PEDE+$w_{\rm dm}$, a strong evidence showing the data set CMB+BAO favor $\Lambda$CDM over PEDE+$w_{\rm dm}$, and a very strong evidence showing the data set CMB+BAO+Pantheon favor $\Lambda$CDM over PEDE+$w_{\rm dm}$. In addition, the Bayesian evidence shows that there are only two data sets, i.e. CMB+Pantheon and CMB+BAO+Pantheon, favor PEDE+$w_{\rm dm}$ over PEDE. To be more specific, there is a strong evidence showing the data set CMB+Pantheon favor PEDE+$w_{\rm dm}$ over PEDE and a positive evidence showing the data set CMB+BAO+Pantheon favor PEDE+$w_{\rm dm}$ over PEDE. As for other two data sets, they all favor PEDE over PEDE+$w_{\rm dm}$ in a positive level.

\section{concluding remarks}
It is well known that there are several long-standing problems implying the discordance of the $\Lambda$CDM model. Although most of solutions trying to resolve these problems assume that the dark matter is cold, astronomical observations from various sources have yet to dismiss the idea that dark matter could be non-cold. In other words, the nature of dark matter remains unknown, therefore, it is natural to treat the dark matter equation of state parameter as a free parameter, and employ observational data to verify whether it is zero, which is what we have done in this article. The PEDE model that shares the same parameter space with $\Lambda$CDM was proposed by the authors of Ref.~\cite{li2019simple} for the purpose to resolve the Hubble tension. In order to check how the extra degrees of freedom in terms of the neutrino properties could affect the constraints on the Hubble constant, the PEDE+$M_{\nu}+N_{\rm eff}$ model is considered by the authors of Ref.~\cite{yang2021emergent}, with the full latest data set combination, they found an indication for non-zero neutrino mass with a significance above 2$\sigma$. This leads us to speculate that it's possible that the non-zero dark matter equation of state is favored by current observations with a significance exceeding 2$\sigma$ when PEDE plays the role of dark energy, therefore, we propose the PEDE+$w_{\rm dm}$ model.

Considering the evolution of the PEDE+$w_{\rm dm}$ model both at the background and perturbation levels, we present some analysis concerning the impacts of the PEDE+$w_{\rm dm}$ model on the CMB TT and matter power spectra for different values of $w_{\rm dm}$, and find that negative values of $w_{\rm dm}$ result in many changes on the spectra that can be recognized visually, on the one hand, the amplitudes of the peaks of the CMB are increased and the positions of the peaks are moved to the left side, one the other hand, the CMB TT power spectrum is depressed at large scale. In addition to that, negative values of $w_{\rm dm}$ decrease the matter power spectrum. We have also fitted the model using data sets consisting of CMB, CMB+BAO, CMB+Pantheon, and CMB+BAO+Pantheon. We find that the value of parameter $w_{\rm dm}$ is negative at $95\%$ CL for CMB+Pantheon, and CMB+BAO+Pantheon data sets, this implies that we should seriously consider whether 'cold dark matter' is indeed cold. It is worth noting that even by allowing a negative $w_{\rm dm}$, the $H_0$ tension still can be relieved to 1.2$\sigma$ (CMB), 2.0$\sigma$ (CMB+BAO), 3.8$\sigma$ (CMB+Pantheon), and 2.8$\sigma$ (CMB+BAO+Pantheon), expect the 3.8$\sigma$ tension for CMB+Pantheon data set, these levels of tensions all can be attributed to the statistical fluctuations.
It is also noteworthy that, although PEDE+$w_{\rm dm}$ outperforms $\Lambda$CDM in addressing Hubble tension for the majority of datasets, Bayesian evidence suggests that $\Lambda$CDM is favored more by all the datasets considered in this study. In addition, Bayesian evidence also suggests that PEDE is favored over PEDE+$w_{\rm dm}$ by CMB, and CMB+BAO data set.
\section*{Acknowledgments}
This work is supported by the National key R\&D Program of China (Grant No. 2020YFC2201600), National Natural Science Foundation of China (NSFC) under Grant No. 12073088, Guangdong Major Project of Basic and Applied Basic Research (Grant No.2019B030302001), and National SKA Program of China No. 2020SKA0110402.

\bibliographystyle{spphys}
\bibliography{wDM}

\end{document}